\documentclass[onecolumn,showpacs,floatfix,11pt,nofootinbib]{revtex4}
\usepackage{amsfonts}
\usepackage{eurosym}
\usepackage{amssymb,amsmath}
\usepackage{graphicx,float}
\usepackage{indentfirst}
\usepackage{hyperref}

\newcommand{\be}{\begin{equation}}
\newcommand{\ee}{\end{equation}}

\newcommand{\ba}{\begin{eqnarray}}
\newcommand{\ea}{\end{eqnarray}}

\begin{document}

\title{Aspects of semilocal BPS vortex in systems with Lorentz symmetry
breaking}
\author{C. H. Coronado Villalobos}
\email{ccoronado@feg.unesp.br}
\author{J. M. Hoff da Silva}
\email{hoff@feg.unesp.br}
\author{M. B. Hott}
\email{hott@feg.unesp.br}
\affiliation{Departamento de F\'{\i}sica e Qu\'{\i}mica, UNESP - Univ Estadual Paulista,
Av. Dr. Ariberto Pereira da Cunha, 333, Guaratinguet\'{a}, SP, Brazil}
\author{H. Belich}
\email{belichjr@gmail.com}
\affiliation{Departamento de F\'{\i}sica e Qu\'{\i}mica, Universidade Federal do Esp\'{\i}
rito Santo (UFES), Av. Fernando Ferrari, 514, 29060-900, Vit\'oria, ES,
Brazil}

\begin{abstract}
It is shown the existence of a static self-dual semilocal vortex
configuration for the Maxwell-Higgs system with a Lorentz-violating CPT-even
term. The dependence of the vorticity upper limit on the Lorentz-break term
is also investigated.
\end{abstract}

\maketitle

%\flushbottom

\section{Introduction}

The Standard Model (SM) has recently passed its final test. The discovery of
the Higgs boson has confirmed the last prediction of a model of undisputed
success. Despite the tremendous success of this model, it presents a
description of massless neutrinos, and cannot incorporate gravity as a
fundamental interaction.

We expect that a new physics may appear if we reach the TeV scale and
beyond. But if General Relativity and SM are effective theories, which could
be guide concepts to obtain a physics beyond SM? The Higgs mechanism is a
fundamental ingredient, used in the electroweak unification, to obtain the
properties of low-energies physics. The breaking of a symmetry by a scalar
field describing a phase transition is currently used in many branches of
sciences. Without going into details, we would say that at the microscopic
level an effective field generated spontaneously can give clues on how to
get the fundamental theory. In relativistic systems, the field that realizes
the breaking must be a scalar in order to preserve Lorentz symmetry.

In nonrelativistic quantum systems, phase transitions such as in
ferromagnetic systems, the rotation symmetry is broken due to the influence
of a magnetic field. For relativistic systems, the realization of symmetry
breaking can be extended by considering a background given by a constant $4$%
-vector field that breaks the symmetry $SO\left( 1,3\right) $ and no longer
the symmetry $SO\left( 3\right) $. This new possibility of spontaneous
violation was first suggested by Kostelecky and Samuel \cite{extra3} in
1989, indicating that, in the string field theory scenario, the spontaneous
violation of symmetry by a scalar field could be extended to other classes
of tensor fields.

This line of research including spontaneous violation of the Lorentz
symmetry in the Standard Model is known in the literature as Standard Model
Extension (SME) \cite{PRIN, povo}, and the breaking is implemented by
condensation of tensors of rank $>1$. This program includes investigations
over all the sectors of the standard model --- fermion, gauge, and Higgs
sectors (a very incomplete list includes \cite{LISTA}) --- as well as
gravity extensions \cite{GRAV}. Following this reasoning, the study of
topological defects has also entering this framework \cite{TDQ}. Quite
recently \cite{QR}, it was demonstrated that a Maxwell-Higgs systems with a
CPT-even Lorentz symmetry violating term yields
Bogomol'nyi-Prasad-Sommerfield (BPS) \cite{BPS} vortex solutions enjoying
fractional quantization of the magnetic field.

Topological defects arising from the spontaneous symmetry breaking are
physical systems of interest in a wide range of theories, from condensed
matter to cosmology \cite{LIV}. These defects may arise from an abelian, as
well as non-abelian, spontaneously broken symmetry. The type of the defect
depends on the broken symmetry. Among the typical interesting defects,
vortex solutions are a relevant class and their characteristics have been
extensively investigated in the literature \cite{SORT}. So, an interesting
program would be to investigate topological defects in a scenario with the
violation of Lorentz symmetry and to identify all those quantities which can
be directly affected by this special type of breaking, namely
Lorentz-symmetry breaking.

One of the benchmarks of the vortex theory is the semilocal vortex \cite{VA}%
. Usually, most part of the study of vortex was restricted to the local
symmetry. However, the inclusion of a global symmetry, besides the usual
local one, may lead to some interesting characteristics in the resulting
topological defect as the presence of topological vortex even if the vacuum
manifold is simply connected, the presence of infinite defects, and the fact
that semilocal strings may end in a cloud of energy.

This paper is partially concerned with the demonstration that semilocal
vortices may be found in a usual Maxwell-Higgs system plus a CPT-even
Lorentz symmetry violating term. In other words, it is possible to combine
the generalized vortex solutions found in \cite{QR} and the semilocal
structure (Section II). As it is well known from the standard properties of
the semilocal setup, the minimum of the potential is a three-sphere, which
is simply connected. In fact, starting from a $SU(2)_{global}\otimes
U(1)_{local}$ symmetry, the symmetry breaks down to $U(1)_{local}$. Hence,
the first homotopy group is trivial, i. e., $\pi _{1}(SU(2)_{global}\otimes
U(1)_{local}/U(1)_{local})=1$. However, the local symmetry also plays its
role. At each point on the three-sphere the local symmetry engenders a
circle. In this vein, looking at the local symmetry, one realizes that it is
possible to obtain infinitely many vortex solutions, corresponding to the
breaking of the local symmetry ($\pi _{1}(U(1)_{local}/1)=\mathbb{Z}$).
Since the potential we shall deal with goes as usual, it is possible to say
that as in the usual Higgs-Maxwell case \cite{VA} , when no
Lorentz-violating term is present, the arguments in favor of stable vortices
are strong, but not exhaustive. In order to guarantee the existence of
semilocal vortices in the Maxwell-Higgs plus Lorentz-violating model, we
have to construct the solutions.

It was shown \cite{QR} that the presence of the Lorentz symmetry violating
term may lead to a peculiar effect in the vortex size. Hence, in view of the
aforementioned characteristic of the semilocal vortex, the solution
combining both effects may result in a most malleable defect structure,
which is shown to be the case. Besides, we show that the Bradlow's limit 
\cite{BRA,DU} depends on the magnitude of a parameter related to the
Lorentz-breaking term, i. e., the vorticity is also affected. In fact, the
vorticity increases as the Lorentz violating term becomes more relevant.

\section{Semilocal vortex with a Lorentz symmetry-breaking term}

We start from the lagrangian density 
\begin{equation}
\mathcal{L}=-\frac{1}{4}F_{\mu \nu }F^{\mu \nu }-\frac{1}{4}(\kappa
_{F})_{\mu \gamma \alpha \beta }F^{\mu \gamma }F^{\alpha \beta }+|D_{\mu
}\Phi |^{2}-\frac{\lambda ^{2}}{4}(\eta ^{2}-|\Phi |^{2})^{2},  \label{1}
\end{equation}
where $\Phi $ is given by the $SU(2)$ doublet $\Phi^{T} =(\phi \,\, \psi)$.
The covariant derivative is given by $D_{\mu }=\partial _{\mu }-ieA_{\mu }$
and $F_{\mu \nu }$ is the usual electromagnetic field strength, in such a
way that the above lagrangian is endowed with the $SU(2)_{global}\otimes
U(1)_{local}$ symmetry. Note that it is similar to the lagrangian
investigated in \cite{QR}, except by the presence of the global symmetry.

The $(\kappa _{F})_{\mu \gamma \alpha \beta }$ term is the CPT-even tensor.
It has the same symmetries as the Riemann tensor, plus a constraint coming
from double null trace $(\kappa _{F})_{\;\;\;\;\alpha \beta }^{\alpha \beta
}=0$. It may be defined according to 
\begin{equation}
(\kappa _{F})^{\mu \gamma \alpha \beta }=\frac{1}{2}(\eta ^{\mu \alpha
}\kappa ^{\gamma \beta }-\eta ^{\gamma \alpha }\kappa ^{\mu \beta }+\eta
^{\gamma \beta }\kappa ^{\mu \alpha }-\eta ^{\mu \beta }\kappa ^{\gamma
\alpha }),  \label{2}
\end{equation}
from which it is readily verified that 
\begin{equation}
(\kappa _{F})_{\mu \gamma \alpha \beta }F^{\mu \gamma }F^{\alpha \beta
}=2\kappa _{\mu \beta }F_{\alpha }^{\mu }F^{\alpha \beta }.  \label{3}
\end{equation}

As we want to generalize the uncharged vortex solution, it is necessary to
set $\kappa _{0i}=0$, since from the stationary Gauss law obtained from (\ref%
{1}) this last condition decouples the electric and magnetic sectors. Hence,
considering the temporal gauge $A_{0}=0$, the energy functional is given by 
\begin{equation}
E=\int d^{2}x\Bigg[\frac{1}{2}[(1-tr(\kappa _{ij}))\delta _{ab}+\kappa
_{ab}]B_{a}B_{b}+|\overrightarrow{D}\Phi |^{2}+\frac{\lambda ^{2}}{4}(\eta
^{2}-|\Phi |^{2})^{2}\Bigg].  \label{4}
\end{equation}

By working with cylindrical coordinates from now on, we implement the
standard vortex ansatz 
\begin{eqnarray}
\phi &=&\eta g_{1}(r)e^{in\theta },  \notag \\
\psi &=&\eta g_{2}(r)e^{in\theta _{2}},  \label{5} \\
A_{\theta } &=&-\frac{1}{er}[a(r)-n].  \notag
\end{eqnarray}%
. The functions $g_{1}(r)$, $g_{2}(r)$, are regular functions and in the
case of a typical vortex solution they have no dynamics as $r\rightarrow
\infty $. It is quite enough to assure that the coupling of the fields to
the gauge field leads to the phases correlation $\theta _{2}=\theta +c$,
being $c$ a constant. Obviously, for a typical vortex solution we shall have
the following boundary conditions for $a(r)$ 
\begin{equation}
a(r)\rightarrow n~\mathrm{as~}r\rightarrow 0~\mathrm{and~}a(r)\rightarrow 0 
\mathrm{as} r\rightarrow \infty  \label{6}
\end{equation}%
. With the chosen ansatz, the magnetic field is trivially given by 
\begin{equation}
B_{z}\equiv B=-\frac{1}{er}\frac{da}{dr}.  \label{8}
\end{equation}%
. Now, taking $\kappa _{11}+\kappa _{22}=s$ it is possible to write 
\begin{equation}
E=\int d^{2}r\Bigg[\frac{1}{2}(1-s)\frac{1}{e^{2}r^{2}}\Big(\frac{da}{dr} %
\Big)^{2}+\eta ^{2}(\frac{dg_{1}}{dr}\Big)^{2}+\eta ^{2}(\frac{dg_{2}}{dr} %
\Big)^{2}+\frac{a^{2}\eta ^{2}}{r^{2}}(g_{1}^{2}+g_{2}^{2})+\frac{\eta
^{4}\lambda ^{2}}{4}(1-g_{1}^{2}-g_{2}^{2})^{2}\Bigg].  \label{9}
\end{equation}%
. By imposing the self-duality condition \cite{QR} $\lambda ^{2}=2e^{2}/(1-s)
$ --- the equivalent to the equality of the scalar and gauge field masses
--- it is possible to rearrange the terms in (\ref{9}) after a bit of
algebra, such as 
\begin{eqnarray}
E &=&\left. \int d^{2}r\Bigg[\frac{(1-s)}{2}\Bigg(\frac{1}{er}\frac{da}{dr}
\pm \frac{e\eta ^{2}}{(1-s)}(1-g_{1}^{2}-g_{2}^{2})\Bigg)^{2}+\eta ^{2}%
\Bigg( \mp \frac{ag_{1}}{r}+\frac{dg_{1}}{dr}\Bigg)^{2}+\eta ^{2}\Bigg(\mp 
\frac{ ag_{2}}{r}+\frac{dg_{2}}{dr}\Bigg)^{2}\right.  \notag \\
&\pm &\left. \frac{\eta ^{2}}{r}\Bigg(\frac{d(ag_{1}^{2})}{dr}\Bigg)\pm 
\frac{\eta ^{2}}{r}\Bigg(\frac{d(ag_{2}^{2})}{dr}\Bigg)\mp \frac{\eta ^{2}}{r%
}\frac{da}{dr}\Bigg].\right.  \label{10}
\end{eqnarray}

In the above expression, the linear terms are those which contribute to the
minimum energy when the self-dual equations are fulfilled. The first order
BPS equations are given by 
\begin{equation}
\frac{dg_{1}}{dr}=\pm \frac{g_{1}a}{r},\frac{dg_{2}}{dr}=\pm \frac{g_{2}a}{r}
\label{11}
\end{equation}
and 
\begin{equation}
-\frac{1}{er}\frac{da}{dr}=\pm \frac{e\eta ^{2}}{(1-s)}
(1-g_{1}^{2}-g_{2}^{2}),  \label{13}
\end{equation}
while the energy minimum is given by 
\begin{equation}
E_{min}=\pm 2\pi \eta ^{2}n(1-g_{1}^{2}(0)-g_{2}^{2}(0)),  \label{14}
\end{equation}
where in the last equation we have used the boundary conditions (\ref{6})
and the fact that $g_{1}$ and $g_{2}$ are regular functions.

Now we are in position to show that despite the fact that the vacuum
manifold is simply connected the field configuration vanishing at the center
of the vortex is compatible with the above framework. Introducing $%
g^{2}=g_{1}^{2}+g_{2}^{2}$, subject to the boundary condition $g\rightarrow
0 $ as $r\rightarrow 0$, one immediately gets 
\begin{equation}
E_{\min }=\pm 2\pi \eta ^{2}n=\eta ^{2}e|\Phi _{B}|,  \label{15}
\end{equation}
where $\Phi _{B}$ is the magnetic flux, and 
\begin{equation}
B=-\frac{1}{er}\frac{da}{dr}=\pm \frac{e\eta ^{2}}{(1-s)}(1-g^{2}).
\label{16}
\end{equation}
The two remaining equations may be bound together as 
\begin{equation}
\frac{dg}{dr}=\pm \frac{ga}{r}.  \label{18}
\end{equation}
Equations (\ref{16}) and (\ref{18}) are identical to the self-dual equations
found in \cite{QR}. Therefore the same conclusions obtained there are
applicable to the present semilocal case. Of particular interest, their
numerical results attest the stability of the BPS vortex solutions.

At this point it would be interesting to say a few words concerning the
semilocal solutions. From the first order equations (\ref{11}) and (\ref{18}
), it is easy to see that 
\begin{equation}
\frac{1}{g}\frac{dg}{dr}=\frac{1}{g_i}\frac{dg_i}{dr},  \label{disc1}
\end{equation}
where i=1,2. Hence the solutions shall obey $g\sim g_i$ and by the
constraint $g^2=g_1^2+g_2^2$ we have $1=f_1^2+f_2^2$, where $f_i$ are
numerical factors. Thus, we see that there are plenty configurations
satisfying the boundary conditions. Each of this configurations corresponds
to a local spontaneous symmetry breaking of the vacuum manifold. In fact,
the vacuum manifold associated to the $SU(2)_{global}\times U(1)_{local}$
symmetry may be understood as a three-sphere whose \textit{each point} (due
to the \textit{local} symmetry) is given by a $S^1$ circle. It is nothing
but the fiber bundle formulation of the vacuum, being the base space that
one associates to the $SU(2)_{global}$ (the three-sphere) and the typical
fibre performed by the manifold associated to the $U(1)_{local}$ ($S^{1}$
circles). The projections are global transformations while the fibre is a
gauge transformation. A particular solution of $(1=f_1^2+f_2^2)$ means a
given $S^1 \rightarrow S^1$ mapping performed by $\Phi$. The infinitely many
possibilities evinced by the equation $1=f_1^2+f_2^2$ stands for the
infinite possibilities of local symmetry breaking, see Figure 1. Finally,
the situation from the vacuum manifold is clear: the local symmetry breaking
lead to special vortex configurations which can end since the base manifold
is simply connected.

\begin{figure}[]
\begin{center}
\includegraphics[width=8in]{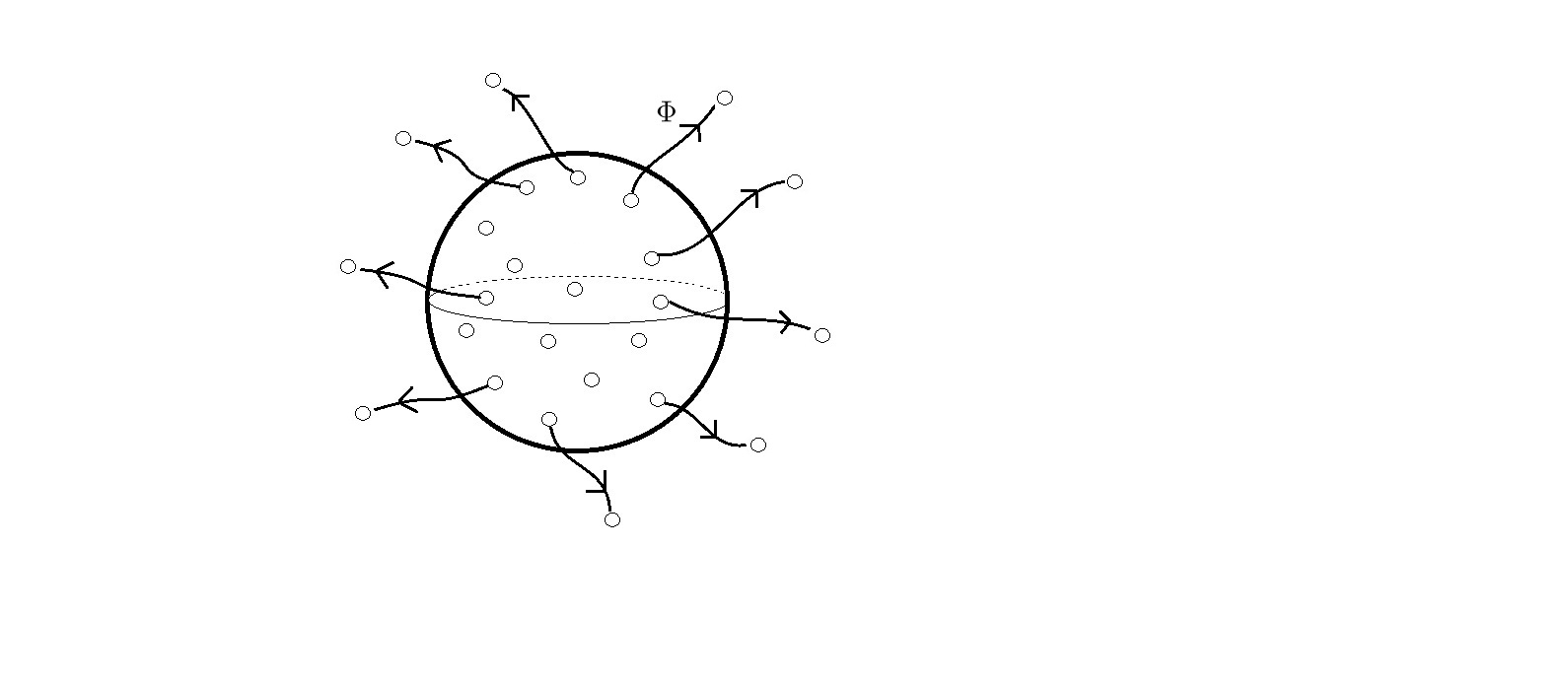}
\end{center}
\caption{{\protect\small The vacuum manifold: the infinitely many
spontaneous local symmetry breaking and the mapping in vortex configurations.%
}}
\end{figure}

\section{$s$ parameter and Bradlow's limit}

It was demonstrated in \cite{QR} that the Lorentz-violating parameter $s$
plays an important role acting as an element able to control both the radial
extension and the amplitude of the defect. In summary, the bigger the $s$
parameter the more compact is the vortex in the sense that the scalar field
reaches its vacuum value (or, equivalently, the gauge field goes to zero) in
a reduced radial distance in comparison with the situation when the Lorentz
symmetry is preserved, the so-called Abrikosov-Nielsen-Olesen vortices.
Thus, if one applies this model to the scenario of type-II superconductors
one sees that the Lorentz symmetry-breaking term is responsible to enhance
the superconducting phase.

It is insightful to relate this effect with the maximum vorticity which a
noninteracting static vortex system may acquire in a given compact base
manifold of area $\mathcal{A}$. This upper bound on the vorticity is the
so-called Bradlow's limit \cite{BRA}. Integrating over equation (\ref{16}),
and choosing positive vorticity, it is easy to see that 
\begin{equation}
\frac{1}{e}\int_{0}^{2\pi }d\theta \int_{0}^{\infty }rdr\frac{1}{r}\frac{da}{
dr}=\int d^{2}r\frac{e\eta ^{2}}{1-s}(1-g^{2}),  \label{19}
\end{equation}
and then 
\begin{equation}
n\leq \frac{e^{2}\eta ^{2}}{2\pi (1-s)}\mathcal{A},  \label{20}
\end{equation}
Note that for $s=0$ the usual Bradlow's limit is recovered, as expected. As $%
s$ grows, however, also does the upper limit. In other words it is possible
to saturate the manifold with more vorticity.

If we contrast this situation with the information that as $s$ grows the
vortex become more compact, we see that these two effects are related: the
increasing $s$ the more compact the vortex. The more compact the vortex,
more vortices with vorticity one are allowed within the same base manifold.
This may be regarded to the growth in the number of vortices shown in a
condensed matter vortex sample under an external (fixed direction) magnetic
field, or the reduction of the vortex core size due to an increase in the
rotation frequency of an electrically neutral superfluid.

Finally, as $s$ approaches $1$ it is possible to see that the Bradlow's
limit blows up. Again, it is in consonance with the analysis performed in 
\cite{QR}, where this limit means an extremely short-range theory in which
the vortex core length goes to zero but the intensity of the magnetic inside
the vortex increases. Similarly to what happens in type II-superconductors,
a phase transition where the multiplicity of vortices with vorticity one is
favored rather than the melting of the condensate might occur. Such behavior
of the magnetic field reinforcing the superconducting phase occurs in
ferromagnetic materials that have coexisting ferromagnetism with
superconductivity \cite{DIMO}. Parenthetically, if one wants to be in touch
with quantum field theory bounds, we notice that the bounds $s\in (-1,\sqrt{%
2 }-1)$ can be obtained by comparison with the results found in the detailed
study carried out in \cite{BEL} for the bounds on the parameter in order to
guarantee not only the causality and the unitarity in the dynamic regime,
but also the stability of vortex-like configurations (stationary regime) in
the Abelian-Higgs model with the CPT-even Lorentz symmetry term in the
electromagnetic sector. If we are interested in preserve causality but relax
the unitarity of the model, we have to take into account the whole interval $%
s\in (-1,1)$ (we obtain this domain using the results of ref. \cite{BEL}
which was adapted to our case). In verifying possible instabilities, such as
those resulting in phase transitions, one has to consider the values of $s$
in this range. As $s$ approaches 1 it is possible to see that the Bradlow's
limit blows up, signalizing a phase transition.

On the other hand, it is expected that the spontaneous violation of Lorentz
symmetry occurs at high-energy (Grand Unification or Planck scale), while in
our energy scale it is manifested only very weakly. In \cite{klink2} the
same violating term used in our article is investigated with $0<s \ll 1$. In
fact, the reference \cite{tab} presents a table with possible values of
Lorentz-symmetry breaking parameters for wide class of violating sources in
the context of the SME. By considering the results presented in \cite{tab}
when the even sector of the SME is taken into account we conclude that $%
\left\vert s\right\vert \,<\,10^{-14}$. Then, for those allowed values of $s$%
, we can not see this transition, once the phase transition might occur for $%
s \rightarrow 1$.

\section{Final remarks}

Is was shown the existence of semilocal BPS vortices in the Maxwell-Higgs
model with a Lorentz symmetry-breaking CPT-even term. The model has,
initially, the $SU(2)_{global}\otimes U(1)_{local}$ and it was demonstrated
that minimum energy configurations are found when the scalar field doublet
vanishes at the center of the core, even being its vacuum manifold simply
connected. As in \cite{VA}, the vacuum manifold may be understood as a
three-sphere pierced by $S^{1}$ circles at each point. Hence, there are
infinitely many vortices appearing in the local breaking $%
U(1)_{local}\rightarrow 1$. These configurations correspond to the (also
infinitely many) possibilities that $g$ may achieve its boundary conditions
(remember that $g^{2}=g_{1}^{2}+g_{2}^{2}$). Going further we studied the
effects of the Lorentz symmetry-breaking term on the vorticity, relating it
with the analysis performed for the usual vortex solution in this type of
system \cite{QR}.

We would like to remark that, although the gauge structure of the model
still retain $SU_{global}(2)\otimes U_{local}(1)$ invariance, it is not
evident that the modification of the gauge field kinetic term not
necessarily can be always made without spoiling the solution achievement.
Particularly, in the case treated here we have resort to the constraint $%
g^2=g_1^2+g_2^2$ and to simple algebraic procedures to reach equation (\ref%
{18}) from (\ref{11}). The point to be stressed is that the gauge field
information, encoded in $a(r)$, must be the same for both parts of the
scalar doublet, otherwise the solution cannot be reached. Moreover, among
all the possibilities brought on by the Lorentz symmetry breaking term, the
interesting one, which does not jeopardize the formal construction of the
stable BPS solutions, is given when $\kappa_{0i}$ vanishes, leading to the
functional form of the energy as in (\ref{9}). It turns out that, after all,
this possibility appears to be appealing, since it possesses quantum field
theory boundaries on its magnitude and, as investigated, leads to an
interesting shift in the Bradlow's limit, which can be physically
interpretable.

It may be instructive to point out a counter example. Suppose a Lagrangian
whose Maxwell kinetic term is present, but with another gauge field ruled by
an Abelian Chern-Simons term as well (the gauge potential being, then, a
simple sum of the Maxwell and Chern-Simons standard potentials). The
mathematical structure of the action is the same $SU_{global}(2)\otimes
U_{local}(1)$. Therefore, everything would go as usual. However, if one
retains the same scalar field potential it is not possible to achieve a
solution via the binding procedure, amd one would not have semilocal
vortices.

Usually, the search for stable vortex solutions is restricted to
modifications of the scalar potential when the gauge field sector of the
model is modified. Within this context, every modification leading to an
explicit vortex solution, as well as its semilocal generalization, deserves
attention. We believe that the (even preserving the mathematical form)
modifications on the gauge side of the symmetry are in the same footing as
those in the potential sector, and the solutions must be \textit{explicitly}
constructed and studied.

\section*{Acknowledgments}

The authors thank to CAPES and CNPq for the financial support. JMHS thanks
to Aristeu Lima for insightful conversation, and the Niels Bohr Institute
where this work was partially done.


\begin{thebibliography}{99}
\bibitem{extra3} {\footnotesize V. A. Kostelecky and S. Samuel, Phys. Rev.
Lett. \textbf{63}, 224 (1989).}

\bibitem{PRIN} {\footnotesize D. Colladay and V. A. Kostelecky, Phys. Rev. D 
\textbf{55}, 6760 (1997); D. Colladay and V. A. Kostelecky, Phys. Rev. D 
\textbf{58}, 116002 (1998). }

\bibitem{povo} {\footnotesize K. Bakke et al., Eur. Phys. J. Plus \textbf{127%
}, 102 (2012); K. Bakke et al., J. Phys. G: Nucl. Part. Phys. \textbf{39},
085001 (2012); K. Bakke et  al., J. Phys. G: Nucl. Part. Phys. \textbf{40},
065002 (2013); K. Bakke et al., Ann. Phys. \textbf{333}, 272 (2013); R.
Casana et al., Phys. Lett. B \textbf{718}, 620 (2012);  R. Casana et al.,
Phys. Rev. D \textbf{86}, 125033 (2012); R. Casana et al., Phys. Lett. B 
\textbf{726}, 488 (2013), R. Casana et al., Phys. Lett. B \textbf{726}, 815
(2013),  R. Casana et al., Phys. Rev. D \textbf{87}, 047701 (2013).}

\bibitem{LISTA} {\footnotesize S.M. Carroll, G.B. Field, and R. Jackiw,
Phys. Rev. D \textbf{41}, 1231 (1990); V. A. Kostelecky and M. Mewes, Phys.
Rev. Lett. \textbf{87}, 251304  (2001); D. Colladay and V. A. Kostelecky,
Phys. Lett. B \textbf{511}, 209 (2001); V.A. Kostelecky and C. D. Lane, J.
Math. Phys. \textbf{40}, 6245 (1999); M. Gomes, J. R.  Nascimento, A. Yu.
Petrov, and A. J. da Silva, Phys. Rev. D \textbf{81}, 045018 (2010); F. A.
Brito, E. Passos, and P.V. Santos, Europhys. Lett. \textbf{95}, 51001
(2011); K.  Bakke, H. Belich, and E. O. Silva, J. Math. Phys. \textbf{52},
063505 (2011); J. L. Boldo, J. A. Helayel-Neto, L. M. de Moraes, C. A. G.
Sasaki and V. J. V. Otoya, Phys. Lett. B  \textbf{689}, 112 (2010). }

\bibitem{GRAV} {\footnotesize V. A. Kostelecky and J. D. Tasson, Phys. Rev.
D \textbf{83}, 016013 (2011). }

\bibitem{TDQ} {\footnotesize M. D. Seifert, Phys. Rev. D \textbf{82}, 125015
(2010); A. de Souza Dutra, M. Hott, and F. A. Barone, Phys. Rev. D \textbf{74%
}, 085030 (2006); D.  Bazeia, M. M. Ferreira Jr., A. R. Gomes, and R.
Menezes, Physica D (Amsterdam) \textbf{239}, 942 (2010); A. de Souza Dutra
and R. A. C. Correa, Phys. Rev. D \textbf{83}, 105007  (2011); A. P. Baeta
Scarpelli and J. A. Helayel-Neto, Phys. Rev. D \textbf{73}, 105020 (2006). }

\bibitem{QR} {\footnotesize C. Miller, R. Casana, M. M. Ferreira Jr., and E.
da Hora, Phys. Rev. D \textbf{86}, 065011 (2012). }

\bibitem{BPS} {\footnotesize E. B. Bogomol'nyi, Sov. J. Nucl. Phys. \textbf{%
\ 24}, 449 (1976); M. K. Prasad and C M. Sommerfield, Phys. Rev. Lett. 
\textbf{\ 35}, 760 (1975). }

\bibitem{LIV} {\footnotesize M. I. Monastyrsky, \textit{Topology of Gauge
Fields and Condensed Matter}, Plenum Press (1993); A. Vilenkin and E. P. S.
Shellard, \textit{Cosmic strings and other topological defects}, Cambridge
University Press (2000); T. Vachaspati, \textit{Kinks and domain walls: an
introduction to classical and quantum solitons}, Cambridge  University Press
(2006).}

\bibitem{SORT} {\footnotesize A. Vilenkin, Phys. Rep. \textbf{121}, 263
(1985).}

\bibitem{VA} {\footnotesize T. Vachaspati and A. Ach\'ucarro, Phys. Rev. D 
\textbf{44}, 3067 (1991). }

\bibitem{BRA} {\footnotesize S. B. Bradlow, Commun. Math. Phys. \textbf{135}
, 1 (1990). }

\bibitem{DU} {\footnotesize G. V. Dunne, \textit{Aspects of Chern-Simons
Theory}, Lectures at the 1998 Les Houches Summer School: Topological Aspects
of Low Dimensional Systems,  [arXiv:hep-th/9902115]. }

\bibitem{DIMO} {\footnotesize D.V. Shopova, D.I. Uzunov Phys. Lett. A 
\textbf{313}, 139 (2003).}

\bibitem{BEL} {\footnotesize H. Belich, F. J. L. Leal, H. L. C. Louzada and
M. T. D. Orlando, Phys. Rev. D \textbf{86}, 125037 (2012).}

\bibitem{klink2} {\footnotesize G. Betschart, E. Kant, F.R. Klinkhamer,
Nucl. Phys. B \textbf{815}, 198 (2009).}

\bibitem{tab} {\footnotesize \ V. A. Kostelecky and Neil Russell, Rev. Mod.
Phys. \textbf{83}, 11 (2011).}

\bibitem{valor} {\footnotesize \ R. Casana, et al., Phys. Rev. D \textbf{87}%
, 047701 (2013); A .G. de Lima et al., Eur. Phys. J. Plus \textbf{128}, 154
(2013).}
\end{thebibliography}
\end{document}